\documentclass[12pt]{article}
\usepackage{amsmath,epsfig,graphicx,amssymb}
\usepackage[margin=0.7in]{geometry}
\date{}
\begin{document}
\title{Quantum-Classical Complementarity and Relativistic Massless Free Fields}
\author{Partha Ghose\footnote{partha.ghose@gmail.com} \\
The National Academy of Sciences, India,\\ 5 Lajpatrai Road, Allahabad 211002, India}
\maketitle
\begin{abstract}
It is shown that relativistic wave equations for free, massless fields display quantum-classical complementarity. 
\end{abstract}
\section{Introduction}
The superposition principle distinguishes quantum from classical mechanics but not quantum from classical optics, both of which show coherence and interference effects. Relativistic massless free fields are special in this respect. In this paper it will be shown that a {\em relativistic} Schr\"{o}dinger-like equation (RSE) follows directly from a homogeneous massless wave equation. The demonstration will be given for a monochromatic and stationary wave, and RSE will be shown to reduce to the Helmholtz equation. Alternative mathematical forms of solutions of RSE will be shown to describe quantum and classical waves, displaying {\em quantum-classical complementarity}. This will be shown to be true of both electrodynamics and weak gravitational waves. 

It will also be shown that RSE can be cast in a manifestly Lorentz covariant form, and that the restriction to stationary and monochromatic wave functions is not a serious limitation.
\section{Electrodynamics}
It is well known that complex scalar fields rigorously describe free classical electromagnetic fields \cite{green, wolf}. They satisfy the Lorentz invariant wave equation
\begin{equation}
\left(\partial^2_{\vec{x}} - \frac{1}{c^2}\partial_t^2\right)\psi(\vec{x}, t) = 0. \label{KG}
\end{equation}
Consider a stationary and monochromatic solution of the form
\begin{equation}
\psi(\vec{x}, t) = \chi(\vec{x}){\rm exp}(-i\omega t)\label{st}
\end{equation}
with a complex time-independent amplitude $\chi(\vec{x})$. 
Applying the time derivative operator in the wave equation (\ref{KG}) only once on $\psi$, one can derive the equation
\begin{equation}
i\partial_t \psi(\vec{x},t) = -\frac{c^2}{\omega}\partial^2_{\vec{x}} \psi(\vec{x},t).\label{x}
\end{equation}
This equation has the same operator structure as the nonrelativistic Schr\"{o}dinger equation and will be referred to in the following as the {\em relativistic Schr\"{o}dinger-like equation} (RSE).

There exists also a hermitian displament operator $\hat{d}_i = -i\nabla_i$ so that
\begin{equation}
[x_i, \hat{d}_j] = i\delta_{ij}. \label{com}
\end{equation}
Hence, {\em characteristic quantum mechanical features are already present in the classical field theory} although the reduced Planck constant $\hbar$ does not appear anywhere! This is similar to the Weyl equation for massless Dirac particles which does not involve $\hbar$ although it is quantum mechanical \cite{weyl}.

Eqn (\ref{x}) reduces to the Lorentz invariant Helmholtz equation
\begin{eqnarray}
\left(\partial^2_{\vec{x}} + k^2\right)\psi(\vec{x}, t) &=& 0,\,,\,\,k = \frac{\omega}{c} \label{helm}
\end{eqnarray}
on applying the time derivative operator once more on the wave function. This equation is well known to describe stationary and monochromatic classical light of frequency $\omega$.

Now, it might occur to the reader that the Maxwell wave equation as well as the Schr\"{o}dinger equation are already known to reduce to the Helmholtz equation for stationary states, and so, nothing new has been achieved. A little reflection will show that such is not the case. The Helmholtz equation (\ref{helm}) associated with the relativistic wave equation (\ref{KG}) is Lorentz invariant but not the Helmholtz equation associated with the nonrelativistic Schr\"{o}dinger equation. This shows up through the term $k^2 = \omega^2/c^2$ in eqn (\ref{helm}) which transforms like $\omega^2 \sim t^{-2}$ in keeping with Lorentz invariance unlike the corresponding term $k^2= 2m\omega/\hbar$ in the Helmholtz equation associated with the Schr\"{o}dinger equation, which transforms like $\omega \sim t^{-1}$. The latter has therefore no bearing on relativistic massless fields. 

What happens is that the relativistic wave equation (\ref{KG}) reduces to the relativistic Helmhotz equation (\ref{helm}) through the intermediate step of the RSE (\ref{x}), which can be cast in the conventional Schr\"{o}dinger form
\begin{equation}
i\hbar\partial_t \psi(\vec{x},t) = -\frac{\hbar^2}{2m^*}\partial^2_{\vec{x}} \psi(\vec{x},t) = \hat{H}_0\psi(\vec{x},t)\label{x2}
\end{equation}
with $m^* = \hbar\omega/2c^2$ as an `effective mass'. The rest mass is zero, and since $\omega$ transforms like $t^{-1}$, it is truly a Lorentz invariant equation of the Schr\"{o}dinger form. 
One can define $\psi(\vec{x},t)$ as a position probability amplitude by normalizing it:
\begin{equation}
\int \psi^*\psi dV = 1.
\end{equation}
This is the first surprise. 

Now to the second surprise. If one writes a solution of RSE (\ref{x}) in the polar form
\begin{eqnarray}
\psi(\vec{x},t) &=& \sqrt{\rho(\vec{x})}\,{\rm exp}(i\phi(\vec{x},t)), \label{polar}
\end{eqnarray}
where both $\rho$ and $\phi$ are real functions, substitute it in eqn (\ref{x}) and separate the real and imaginary parts, one gets the pair of coupled equations
\begin{eqnarray}
\frac{\partial \phi(\vec{x},t)}{\partial t} + \frac{(\nabla \phi(\vec{x},t))^2}{\omega/c^2} + \mathcal{Q} &=& 0,\nonumber\\\mathcal{Q} &=& -\frac{c^2}{\omega}\,\frac{\nabla^2 \sqrt{\rho(\vec{x})}}{\sqrt{\rho(\vec{x})}},\label{HJ1} \\
\vec{\nabla}. \left(\rho(\vec{x})\vec{\nabla}\phi(\vec{x},t)\right) &=& 0. \label{cont}
\end{eqnarray}
The term $\mathcal{Q}$ couples the phase and the amplitude of the wave and has the same structure as the quantum potential in quantum mechanics. Just as the quantum potential couples the phase and amplitude in nonrelativistic quantum mechanics and is responsible for quantum coherence and interference, $\mathcal{Q}$ is responsible for coherence and interference in optics.

If one wants to construct a Hamilton-Jacobi theory of electrodynamics, one has to introduce the action $S$. This can be done by writing the phase $\phi = S/\hbar$ with $S(\vec{x},t) = W(\vec{x}) - Et$ for stationary states. Then equation (\ref{HJ1}) can be written as 
\begin{eqnarray}
\frac{\partial S(\vec{x},t)}{\partial t} + H &=& -E + H = 0,\\
H &=& \frac{(\nabla W(\vec{x}))^2}{\hbar\omega/c^2} + \hbar\mathcal{Q}. \label{Hamq}
\end{eqnarray}
This is the Hamilton-Jacobi equation in stationary electrodynamics for the action $S$. This also leads to the natural definition of an effective mass $m^* = \hbar\omega/2c^2$. Then
\begin{equation}
Q = \hbar \mathcal{Q} = - \frac{\hbar^2}{2m^*}\,\frac{\nabla^2 \sqrt{\rho(\vec{x})}}{\sqrt{\rho(\vec{x})}}
\end{equation}
has the familiar form of the quantum potential. One can also define $\vec{\nabla} W = \vec{p} = \hbar \vec{k}$. Then
\begin{equation}
H = \hbar\omega + Q \label{H}
\end{equation}
and eqn (\ref{cont}) takes the form
\begin{equation}
\vec{\nabla}. \left(\rho(\vec{x})\vec{p}\right) = 0 \label{poyn}
\end{equation}   
which is the continuity equation for the field momentum if $\rho(\vec{x}) = |\psi(\vec{x})|^2$ is interpreted as the time-independent field intensity at a point. Eqn (\ref{H}) shows that $Q$ can be interpreted as a vacuum or zero-point energy which is purely quantum mechanical.

The probability current density for a $\psi$ in the polar form (\ref{polar}) with $\phi = S/\hbar$ is
\begin{eqnarray}
\vec{j} &=& -\frac{i\hbar}{2m^*}\left[\psi^* \vec{\nabla}\psi -\vec{\nabla}\psi^* \psi\right]\nonumber\\
&=& -\frac{i\hbar}{2m^*}\frac{2i}{\hbar}\left[\rho \vec{\nabla} W\right] = \frac{1}{m^*}\left[\rho \vec{\nabla} W\right].
\end{eqnarray}
This current satisfies the continuity equation
\begin{equation}
\frac{\partial \rho}{\partial t} + \vec{\nabla}. \vec{j} = 0.
\end{equation}
For stationary fields, $\partial \rho/\partial t = 0$ and hence 
\begin{equation}
\vec{\nabla}.\vec{j} = 0.
\end{equation}
When multiplied by the constant total energy $E$, this can be written as
\begin{equation}
\vec{\nabla}. \vec{S} = 0
\end{equation}
where $\vec{S} = E\vec{j}$ is the Poynting vector.
This is the Poynting theorem. This equation is identical with eqn (\ref{poyn}) if $\vec{S} = \rho\vec{p}$. 

The same equations also follow from the Helmholtz equation (\ref{helm}) if the wave function is of the polar form with $S/\hbar$ as the phase. 

However, if the wave function is of the form (\ref{st}) rather than of the polar form (\ref{polar}), RSE (\ref{x}) and the Helmholtz equation (\ref{helm}) have classical solutions. This is the second surprise.

To recapitulate, let us look at the three equations (\ref{KG}), (\ref{x}) and (\ref{helm}),
\begin{eqnarray}
\left(\partial^2_{\vec{x}} - \frac{1}{c^2}\partial_t^2\right)\psi(\vec{x}, t) &=& 0,\nonumber\\
i\partial_t \psi(\vec{x},t) &=& -\frac{c^2}{\omega}\partial^2_{\vec{x}} \psi(\vec{x},t),\nonumber\\
\left(\partial^2_{\vec{x}} + k^2\right)\psi(\vec{x},t) &=& 0
\end{eqnarray}
for a stationary and free monochromatic $\psi(\vec{x}, t)$.
The first and third equations are familiar classical equations. But, as we have seen, the second equation, which is derived from the first and which leads to the third, has a Schr\"{o}dinger-like structure. With a polar form of the wave function with a real amplitude and the phase expressed as $\phi = S/\hbar$, this equation, as well as the Helmholtz equation, gives rise to quantum mechanical solutions. Thus, {\em the normalized polar form of the wave function with a phase $S/\hbar$ is a wave function of a photon} of energy $\hbar\omega$ and an effective momentum $m^*c =\hbar\omega/2c$. 

On the other hand, if the wave function has a complex time-independent amplitude, the solution is classical. 

Classical and quantum light are mutually exclusive forms of light produced by mutually exclusive state preparation methods such as those for producing thermal light or laser beams and those for producing non-classical states such as spontaneous and parametric down-conversion. It turns out that these mutually exclusive forms of light can be described by alternative mathematical forms of the wave function satisfying RSE. One is thus led to {\em quantum-classical complementarity} of light.
\subsection{Manifest Lorentz Covariance}
The choice of stationary states with a clear separation of the spatial and time components might appear to violate Lorentz invariance, particularly because of the occurrence of the Schr\"{o}dinger-like equation RSE (\ref{x}). That this is not the case can be shown by foliating the Minkowski manifold by a continuous set $\{\sigma\}$ of space-like hypersurfaces whose normal at every point is time-like. The whole sequence of these space-like slices is generated by time evolution. Each point $p \in \sigma$ has its own local time $t_p$ so that there is no preferred time frame that defines absolute simultaneity. This is therefore a Lorentz invariant procedure. Lorentz transformations will transform one set of surfaces $\{\sigma\}$ to another set $\{\sigma^\prime\}$ with local times $t_{p^\prime}, p^\prime \in \sigma^\prime$, and one is free to choose any of these foliations. Thus, space-time separation can be done fully relativistically, and a scalar field can be defined as a functional $\psi[\sigma]$. This foliation of the Minkowski manifold was first used by Tomonaga \cite{tom} and Schwinger \cite{sch} to formulate quantum electrodynamics in a manifestly Lorentz covariant form.

In the interaction or Dirac representation the standard Schr\"{o}dinger equation takes the form
\begin{equation}
i\hbar \partial_t \psi_I(t) = \hat{V}_{I}(t)\psi_I(t)
\end{equation}
where the wave function in the interaction representation is $\psi_I(t) = {\rm exp}(i\hat{H}_{0,S}t/\hbar)\psi_S(t)$, $\psi_S(t)$ being the wave function in the Schr\"{o}dinger representation, $\hat{H}_{S} = \hat{H}_{0,S} + \hat{V}_S$ the Hamiltonian in the Schr\"{o}dinger representation, and $\hat{V}_I(t)$ the interaction Hamiltonian in the interaction representation related to $\hat{V}_S$ by
\begin{equation}
\hat{V}_I(t) = e^{i\hat{H}_{0,S}t/\hbar}\hat{V}_Se^{-i\hat{H}_{0,S}t/\hbar}.  
\end{equation}
This can be written in the manifestly Lorentz covariant form
\begin{equation}
\left[-i\hbar\frac{\delta}{\delta\sigma_p} + \hat{V}_{I p}\right]\psi_I[\sigma] =0
\end{equation}
provided $\hat{V}_{I p}$ is Lorentz invariant, where $\frac{\delta}{\delta\sigma_p}$ is a functional derivative. In interaction free cases this implies that $\psi_I$ is $\sigma$-independent, i.e. the same on the entire set of space-like hypersurfaces generated by a future-directed time-like vector field. Since this is generally true of all $\psi_I$ in the interaction representation, and since all three representations are equivalent, this shows that free wave functions are essentially stationary. Hence, the restriction to stationary monochromatic waves adopted in this paper to derive the main results is not a serious limitation.

A complete description of electrodynamics with a scalar wave function is given in Appendix A.

\section{Weak Gravitational Fields}
In this section the similarity between electrodynamics and linearized gravity will be used to show that a quantum mechanical description of gravity waves can also be given. It is now well known that for weak gravitational fields Einstein's equation
\begin{equation}
R_{\mu\nu} = -8\pi G\left(T_{\mu\nu} - \frac{1}{2}g_{\mu\nu} T^\lambda_\lambda\right)
\end{equation}
can be linearized by writing the metric in the form
\begin{equation}
g_{\mu\nu}(x) = \eta_{\mu\nu} + h_{\mu\nu}(x)
\end{equation}
where $\eta_{\mu\nu}$ is the flat Minkowski metric and $h_{\mu\nu} \ll 1$ is a small perturbation on it. To the lowest order in the perturbation the vacuum equation $R_{\mu\nu} =0$ can be written in the linearized form \cite{th, schutz, gw}
\begin{eqnarray}
\Box h_{\mu\nu} = (\partial^2 -\partial^2/\partial t^2)h_{\mu\nu} = 0,\label{a}\\
\partial_\nu h^\nu_\mu(x) - \frac{1}{2}\partial_\mu h^\nu_\nu = 0.\nonumber
\end{eqnarray}
The first equation is the linearized Einstein equation, the second equation is a gauge constraint, and $h^\gamma_\nu \equiv \eta^{\gamma\delta} h_{\delta\nu}$. A general solution is of the form
\begin{equation}
h_{\mu\nu}(x) = \alpha_{\mu\nu}\,e^{ik_\lambda x^\lambda} + \alpha^*_{\mu\nu}\,e^{-ik_\lambda x^\lambda}
\end{equation}
with $k_\mu^\mu = 0$, $k^\mu = \eta^{\mu\nu}k_\nu$, where $\alpha_{\mu\nu}$ is the polarization tensor which is symmetric in $(\mu,\nu)$ and so has ten components. This number can be reduced to two by making use of Bianchi identities and fixing the gauge. Hence, for a plane wave propagating in the $z$ direction with a fixed frequency $\omega$ and $k^\mu = (\omega,0,0,\omega),\, k.x = \omega(z - t)$ and $c=1$, the general solution can be written as 
\begin{equation}
h_{\mu\nu}(x) = \alpha_{\mu\nu}\,e^{i\omega(z -t)}
\end{equation}
with 
\begin{equation}
\alpha_{\mu\nu} =
\left(\begin{array}{cccc}
 0 & 0 & 0 & 0\\
0 &\alpha_{11} & \alpha_{12} & 0\\
0 & \alpha_{12} & -\alpha_{11} & 0\\
0 & 0 & 0 & 0
\end{array} \right).\label{sol}
\end{equation} 
This shows that there are two independent transverse polarization states with helicity $\pm 2$. The part proportional to $\alpha_{xx} = \alpha_{11}$ is called the {\em plus-polarization} state and is denoted by $+$, and the part proportional to $\alpha_{xy} = \alpha_{12} = \alpha_{21}$ is called the {\em cross-polarization} state and is denoted by $\times$. Hence, it represents a spin-2 field. 

The most general gravitational wave propagating in the $z$ direction is a superposition of waves of fixed $k^\mu$ of the form
\begin{eqnarray}
|h\rangle &=& \left[f_\times(t - z)|\varepsilon_\times\rangle + f_+ (t - z)|\varepsilon_+\rangle\right] = |\psi(t,z)\rangle =|\phi(z)\rangle e^{i\omega(z - t)}\label{hz}
\end{eqnarray}
where
\begin{equation}
|\varepsilon_\times\rangle = \left(\begin{array}{cccc}
 0 & 0 & 0 & 0\\
0 & 0 & 1 & 0\\
0 & 1 & 0 & 0\\
0 & 0 & 0 & 0
\end{array} \right)
\end{equation}
\begin{equation}
|\varepsilon_+\rangle = \left(\begin{array}{cccc}
0 & 0 & 0 & 0\\
0 & 1 & 0 & 0\\
0 & 0 & -1 & 0\\
0 & 0 & 0 & 0
\end{array} \right)
\end{equation}
are the unit polarization tensors and
\begin{equation}
 |\phi(z)\rangle = \left[f_\times(z)|\varepsilon_\times\rangle + f_+ (z)|\varepsilon_+\rangle\right].
\end{equation}
Using the basis vectors $|+\rangle = \left(0,1,0,0\right)^T,\, |-\rangle = \left(0,0,1,0\right)^T$, one can construct $\varepsilon_\times = |+\rangle\otimes|-\rangle + |-\rangle\otimes|+\rangle \equiv |+ -\rangle + |- +\rangle$, and $\varepsilon_+ = |+\rangle\otimes|+\rangle - |-\rangle\otimes|-\rangle \equiv |+ +\rangle - |- -\rangle$. 
The other two basis vectors $\left(1,0,0,0\right)^T$ and $\left(0,0,0,1\right)^T$ are eliminated by choosing the TT gauge which makes the metric perturbation purely spatial and transverse: $h_{tt} = h_{ti} = 0$, $h_i^{\,\,i} = 0$, $\partial_i h_{ij} = 0$. 

It follows that for solutions like (\ref{hz}), eqn (\ref{a}) can be written in the form
\begin{equation}
(\partial^2/\partial t^2 - \partial_z^2)|\psi(t,z)\rangle= 0\label{a2}
\end{equation} 
Operating only the first time derivative on $|\psi\rangle$ in this equation, one obtains
\begin{equation}
i\frac{\partial |\psi(t,z)\rangle}{\partial t} = -\frac{1}{\omega}\partial_z^2 |\psi(t,z)\rangle.\label{i}
\end{equation}
This has the same operator structure as the Schr\"{o}dinger equation. Like in scalar field theory
one can also define an operator $\hat{d}_z = -i\partial_z$ and $\hat{z} = z$ which satisfy the commutation relation
\begin{equation}
[\hat{z}, \hat{d}_z] = i. 
\end{equation}
Hence, {\em these essentially quantum features are already present in the classical weak gravitational wave}. 

Notice that eqn (\ref{i}) can be written in the explicit Schr\"{o}dinger form
\begin{equation}
i\hbar\frac{\partial |\psi(t,z)\rangle}{\partial t} = -\frac{\hbar^2}{2 m^*}\partial_z^2|\psi(t,z)\rangle\label{b}
\end{equation}
with an effective mass $m^* = \hbar \omega/2$. 

Applying the time derivative operator once more on the state results in the Helmhotz equation
\begin{equation}
\left(\partial_z^2 + k^2\right)|\psi(t,z)\rangle = 0 \label{helm2}
\end{equation}
with $k = \omega$. 

Now, defining $\langle \psi| = |\psi\rangle^\dagger$, we have
\begin{equation}
\langle \psi(t,z)| =\langle \phi(z)| e^{-i(kz - \omega t)}. 
\end{equation}
The inner product of two matrices $X$ and $Y$ is defined by the Frobenius product
\begin{equation}
\langle X|Y\rangle = \frac{1}{2}Tr (X^\dagger Y).
\end{equation}
With this definition let us normalize $\psi$ by requiring 
\begin{equation}
\int \langle \psi(t,z)|\psi(t,z)\rangle dV = \int |\phi(z)|^2 dV = 1
\end{equation}
so that $|\phi|$ can be given the interpretation of a probability amplitude in configuration space. 

Now let us see what happens when $\psi$ is written in a polar form. Since $\psi$ is physically a single wave, let us assume it has a single overall phase and that one can write 
\begin{eqnarray}
|\psi(t,z)\rangle &=& R(z) e^{iS(t,z)/\hbar}|\varepsilon\rangle\label{polar2}\\
S(t,z) &=& W(z) - Et,\label{S}
\end{eqnarray}
where $|\varepsilon\rangle =|\varepsilon_+\rangle + |\varepsilon_\times\rangle$, and $R(z), S(z,t)$ are real functions.
Then substituting this form in eqn (\ref{b}) and separating the real and imaginary parts, one gets
\begin{eqnarray}
\left[\frac{\partial S}{\partial t} + \frac{(\partial_z S)^2}{2m^*} + Q\right] &=& 0,\label{HJ}\\
\partial_z \left(R^2\partial_z W(z)\right) &=& 0,\label{contqm}
\end{eqnarray}
where
\begin{equation}
Q = -\frac{\hbar^2 }{2m^*}\,\frac{\partial^2_z R}{R}.
\end{equation}
Using eqn (\ref{S}), eqn (\ref{HJ}) can be written in the form 
\begin{eqnarray}
\frac{\partial S}{\partial t} + H &=& -E + H =0,\\
H &=& \frac{1}{{2m^*}} (\partial_z W)^2 + Q.
\end{eqnarray}
This is the Hamilton-Jacobi equation in linearized gravity if $S$ is identified with the action. Notice that it is $\hbar$ dependent through the term $Q$ although eqn (\ref{a2}) is not. This is because of the use of the polar form (\ref{polar}) of the wave function with a phase $S/\hbar$. We will see that $E = \hbar\omega$ and $p = \hbar k$, justifying the identification of the unit of action with the reduced Planck constant $\hbar$. 

The Schr\"{o}dinger probability current density for a $\psi$ in the polar form (\ref{polar}) is
\begin{eqnarray}
j_z &=& -\frac{i\hbar}{2m^*}\left[\langle \psi|\partial_z\psi\rangle -\langle\partial_z\psi| \psi\rangle\right]\nonumber\\
&=& -\frac{i\hbar}{2m^*}\frac{2i}{\hbar}\left[R^2\partial_z W\right] = \frac{1}{m^*}\left[R^2\partial_z W\right].
\end{eqnarray}
This current satisfies the continuity equation
\begin{equation}
\frac{\partial R^2}{\partial t} + \partial_z j_z = 0.
\end{equation}
For stationary fields, $\partial R^2/\partial t = 0$ and hence 
\begin{equation}
\partial_z j_z = 0.
\end{equation}
When multiplied by the constant total energy $E$, this can be written as
\begin{equation}
\partial_z S_z = 0
\end{equation}
where $S_z = j_z E$ is the gravitational Poynting vector in the $z$ direction.
This is the Poynting theorem in linearized gravity. This equation is identical with eqn (\ref{contqm}) if $\partial_z W = S_z/R^2 = p_z$. If $p_z = \hbar k$, $(\partial_z W)^2/2m^* = \hbar\omega$, and hence $E = \hbar\omega + Q$. Therefore, $Q$ can be interpreted as the vacuum energy which is purely quantum mechanical in nature.

One can check that the expectation values of energy and momentum in the eigenstate $\psi$ in the polar form (\ref{polar}, \ref{S}) are 
\begin{eqnarray}
\langle E\rangle &=& i\hbar\int \psi^* \partial_t\psi dV = \int \hbar\omega R^2 dV = \hbar\omega,\\
\langle p_z\rangle &=& \frac{-i\hbar}{2}\int \psi^* \overleftrightarrow{\partial}_z \psi dV = \int\partial_z W R^2\, dV = p_z = \hbar k.
\end{eqnarray}
This shows that $\psi$ can indeed be consistently interpreted as a wave function of a single graviton of energy $\hbar\omega = h\nu$, momentum $\hbar k = h\nu$ and spin-2.

If, however, the wave function is of the general form (\ref{hz}), eqn (\ref{b}) reduces to the Helmholtz equation (\ref{helm2}), as we have seen, which describes a classical gravitational wave.

Thus, equation (\ref{i}) provides a unified basis for describing both
quantum and classical linearized gravitational fields as complementary aspects of the same field. Quantum mechanical fields require a particular form of solutions for their description, namely the normalized polar form (\ref{polar2}) of $\psi$. Other forms of the solution like eqn (\ref{hz}) describe classical gravitational fields. 

No claim is being made, however, that a general theory of classical and quantum gravity can be constructed.

\section{Concluding Remarks}
Relativistic massless fields are very special in nature, describing, as they do, electrodynamics and gravitational waves (in the weak field approximation). They can both be described by complex scalar fields that satisfy the standard wave equation together with a polarization tensor which describes their spin. The really surprising and interesting thing about them is their hidden quantum mechanical structure revealed by eqns (\ref{x}) and (\ref{i}).
 
What is also equally interesting is that the same set of equations (\ref{KG}, \ref{x}, \ref{helm}) and (\ref{a2}, \ref{i}, \ref{helm2}) describe both classical and quantum mechanical waves depending on whether the solution has a complex time-independent amplitude or the polar form with a real amplitude. This is the mathematical embodiment of the quantum-classical complementarity of light and gravitational waves. The empirical basis of it lies in the mutually exclusive methods of state preparation of classical and non-classical states.

Massless fields also require the introduction of a fundamental unit of action, as explained in Appendix A. For empirical reasons the unit has been chosen to be $\hbar$. However, it drops out of the wave equation, the Schr\"{o}dinger-like equation and the Helmholtz equation. This seems to be true of all massless field equations including the Weyl equation for massless Dirac particles.

By foliating the Minkowski manifold by space-like hypersurfaces and using the interaction representation, the Schr\"{o}dinger equation with a relativistically invariant interaction Hamiltonian can be cast in a manifestly covariant form. It follows from this that all free wave functions are stationary, and hence the restriction to stationary monochromatic waves used in the text above is not a serious limitation. Also, a reference to the classic papers of Tomonaga and Schwinger makes it clear that a multi-particle theory can also be constructed with each particle at a point carrying its own time.

Finally, it must be mentioned that a photon wave function was proposed by Bialynicki-Birula in 1993 \cite{bb} based on the Riemann-Silberstein vector $\vec{F} = \vec{E} + ic\vec{B}$. Although it has some practical value, this wave function does not have all the properties of a Schr\"{o}dinger wave function. Other attempts to introduce the photon wave function involve the use of approximations such as coarse-graining \cite{ina} or a new definition of what one wishes to mean by it \cite{sipe}. Such attempts become unnecessary once a relativistic Schr\"{o}dinger-like equation is in place. Curiously, the Newton-Wigner-Wightman theorem on the nonlocalizability of a massless particle \cite{new, wight} can actually be turned around in support of a localizable photon with an effective mass $\hbar\omega/2c^2$ \cite{ojima}. 

Once a relativistic Schr\"{o}dinger-like equation with an effective mass is shown to exist, a relativistic de Broglie-Bohm theory also becomes possible without recourse to other more complex equations \cite{gh, ghose}.

\section{Acknowledgement}
A somewhat incomplete version of the paper dealing with electrodynamics was presented at the {\em International Conference on Quantum Frontiers and Fundamentals}, Raman Research Institute, Bengaluru, 30th April- 4th May, 2018, and the author is grateful to the organizers for the invitation. He
also thanks the National Academy of Sciences, India for a grant.

\section{Appendix A}
In classical scalar field theory the action is
\begin{equation}
S = \int d^4x \mathcal{L} =- \int d^4x\left[\partial^\mu\psi^*\partial_\mu\psi - \frac{m^2 c^2}{\hbar^2}\psi^*\psi\right]
\end{equation}
and the field $\psi$ has the mechanical dimension $[\psi] = \sqrt{ml}/t$ and $S$ is guaranteed to have the mechanical dimension of `action', i.e. $[S] = ml^2/t$.

For a massless free field theory, however, one has to introduce an independent unit of action $\hbar$ so that one has an effective mass $m^* \propto \hbar\omega/c^2$. In principle the unit of action $\hbar$ is an arbitrary constant which is fixed by comparison with empirical data. 

The correspondence with classical electrodynamics is given by the mapping \cite{green}
\begin{eqnarray}
\mathcal{L} &=& -\partial^\mu\psi^*\partial_\mu\psi = \dot{\psi}^*\dot{\psi} - \vec{\nabla}\psi^* .\vec{\nabla}\psi\nonumber\\
&\mapsto& \frac{1}{c^2}(\vec{\dot{A}})^*.\vec{\dot{A}} - (\vec{\nabla}\times\vec{A})^*.(\vec{\nabla}\times\vec{A})
\end{eqnarray}
where $\vec{A}$ is the vector potential with mechanical dimension $\sqrt{ml}/t$. The Hamiltonian density is given by
\begin{equation}
\mathcal{H} = \pi^*\pi + \vec{\nabla}\psi^* .\vec{\nabla}\psi \mapsto \frac{1}{2}\left[\frac{E^2}{c^2} + B^2 \right]
\end{equation}
and the momentum density by
\begin{equation}
\frac{1}{2 c}(\dot{\psi}^*\vec{\nabla}\psi + \dot{\psi}\vec{\nabla}\psi^*) \mapsto \frac{1}{\mu_0 c^2}(\vec{E}\times\vec{B}). 
\end{equation}

A complete description of an ordinary state in classical electrodynamics can be given more conveniently as the tensor product of the disjoint Hilbert spaces ${\cal{H}}_{path}$ of square integrable scalar functions $\psi$ and a two-dimensional space of polarization states ${\cal{H}}_{pol}$. Hence, a state of unit intensity can be written as $\frac{1}{\sqrt{|\psi|^2}}|\psi\rangle\otimes |\varepsilon\rangle\in {\cal{H}}_{path}\otimes {\cal{H}}_{pol}$ where $\psi$ is a solution of the scalar wave equation (\ref{KG})
and $|\varepsilon\rangle \in {\cal{H}}_{pol}$ is the state 
\begin{equation}
|\varepsilon\rangle = \left(\begin{array}{c} \varepsilon_1\\ \varepsilon_2\end{array}\right) = e^{i\phi}\left(\begin{array}{c}\cos\theta \\ e^{i\chi}\sin\theta\end{array}\right)
\end{equation}
of the transverse polarizations $\varepsilon_1$ and $\varepsilon_2$. In other words, $|\psi\rangle = \psi|\varepsilon\rangle$.

This can also be written as the Jones vector \[|J\rangle = \frac{1}{\sqrt{\langle J|J\rangle}}\left(\begin{array}{c}
 E_x\\ E_y
\end{array} \right) \] with $E_x = \psi_0 \hat{e}_x {\rm exp (i\phi_x)}$ and $E_y = \psi_0 \hat{e}_y {\rm exp (i\phi_y)}$ as the complex transverse electric fields, $\hat{e}_x$ and $\hat{e}_y$ the unit polarization vectors, and $\langle J|J\rangle = \vert E_x\vert^2 + \vert E_y\vert^2 =  |\psi_0|^2$.

\end{document}